\documentclass[a4paper,11pt]{article}
\pdfoutput=1 

\usepackage{jcappub} 

\usepackage[T1]{fontenc} 

\usepackage{units}

\title{\boldmath Forecasting the Interaction in Dark Matter-Dark Energy Models with Standard Sirens From the Einstein Telescope}

\author[a,1]{Riis R. A. Bachega}%
\author[b]{Andr\'e A. Costa}%
\author[a,2]{E. Abdalla}
\author[c, d]{K.S.F. Fornazier}%

\affiliation[a]{Instituto de F\'{i}sica,  Universidade de S\~{a}o Paulo, Caixa Postal 66318, CEP 05314-970, S\~{a}o Paulo, SP, Brazil}
\affiliation[b]{Center for Gravitation and Cosmology, College of Physical Science and Technology, Yangzhou University, Yangzhou 225009, China}
\affiliation[c]{Divis\~{a}o de Astrof\'{i}sica, Instituto Nacional de Pesquisas Espaciais, Avenida dos Astronautas 1758, S\~{a}o Jos\'{e} dos Campos, 12227-010, SP, Brazil}
\affiliation[d]{Department of Physics and Astronomy, University College London}

\emailAdd{rrhavia@if.usp.br}
\emailAdd{andrecosta@yzu.edu.cn}
\emailAdd{eabdalla@usp.br}
\emailAdd{karin.fornazier@gmail.com}

\abstract{Gravitational Waves (GW's) can determine the luminosity distance of the progenitor directly from the amplitude of the wave, without assuming any specific cosmological model. Thus, it can be considered as a standard siren. The coalescence of binary neutron stars (BNS) or neutron star-black hole pair (NSBH) can generate GW's as well as the electromagnetic counterpart, which can be detected in a form of Gamma-Ray Bursts (GRB) and can be used to determine the redshift of the source. Consequently, such a standard siren can be a very useful probe to constrain the cosmological parameters. In this work, we consider an interacting Dark Matter-Dark Energy (DM-DE) model. Assuming some fiducial values for the parameters of our model, we simulate the luminosity distance for a ``realistic" and ``optimistic" GW+GRB events , which can be detected by the third-generation GW detector  Einstein Telescope (ET). Using these simulated events, we perform a Monte Carlo Markov Chain (MCMC) to constrain the DM-DE  coupling constant and other model parameters in $1\sigma$ and $2\sigma$ confidence levels. We also investigate how GW's can improve the constraints obtained by current cosmological probes.}

\begin{document}
\maketitle
\flushbottom

\section{Introduction}
\hspace{0.5 cm}
Observations from type-Ia supernova (SnIa) \citep{Riess:1998cb, Perlmutter:1998np}, Cosmic Microwave Background (CMB) \cite{Ade:2013zuv, Ade:2015xua, Aghanim:2018eyx}, Baryon Acoustic Oscillations (BAO) \cite{Eisenstein:2005su} and Redshift Space Distortions \cite{Kashlinsky:2008ut} have pointed to an acceleration in the expansion of the Universe. This acceleration can be explained by the presence of a negative pressure component, called dark energy (DE).

The standard acceptable model consistent with observations is the $\Lambda$CDM, where the Universe is dominated by cold dark matter (CDM) and  dark energy, which is identified with the cosmological constant $\Lambda$, associated to the vacuum energy with equation of state (EoS) $\omega_{de}=-1$. However, this explanation of DE is not satisfactory from a theoretical point of view, facing the fine tunning \cite{Weinberg:1988cp} and the coincidence problem \cite{Zlatev:1998tr}. The coincidence problem can be stated as follows: how the current values of dark matter and dark energy densities are so similar if the time evolution of each component are so different? To alleviate the coincidence problem, an \textit{interacting dark sector scenario} has been proposed. In such a scenario, energy is exchanged between Dark Matter (DM) and DE. These models are compatible with observations \citep{Abdalla:2014cla, Costa:2016tpb, Marcondes:2016reb}.  Moreover, it has been shown that an interaction in the dark sector can solve the tension in the value of Hubble constant $H_0$ obtained by local and global measurements \cite{Xia:2016vnp, Kumar:2017dnp, DiValentino:2017iww, Yang:2018euj}.  For a more complete review of interacting dark sector models see \citep{Wang:2016lxa}.

In order to better constrain the parameters of each cosmological model, we need to improve the capabilities of current cosmological probes like SnIa, CMB and BAO. Note that all these probes are based on electromagnetic radiation. A different method was first proposed by \citep{Schutz:1986gp} based on gravitational waves (GW) detection from merging compact binaries sources, as binary neutron stars pairs (BNS) or neutron stars-black holes (NSBH). From the gravitational wave signal, we can measure the luminosity distance $d_L$ and, from the electromagnetic counterpart, we can measure the redshift $z$ of the source. Thus, we can construct a $d_L-z$ diagram and constrain the expansion history of the Universe and our cosmological parameters, complementing the current cosmological probes. In analogy to SnIa, which can be considered \textit{standard candles}, GW from merging compact binaries can be considered \textit{standard sirens} (SS). The method is self-calibrating, i.e., do not need any cosmic distance ladder. 

The viability of the standard siren method could only be attested in practice with the first GW detections by LIGO collaboration \citep{Abbott:2016blz}. So far there have been eleven individual detections, ten binary black-holes (BBHs) \citep{Abbott:2016nmj, Abbott:2017vtc, Abbott:2017gyy, Abbott:2017oio, LIGOScientific:2018mvr} and one binary neutron star (BNS) \citep{TheLIGOScientific:2017qsa} during the first and second observation run (O1/O2). The latter is called GW170817 event. The electromagnetic counterpart has also been observed  \citep{Goldstein:2017mmi, Savchenko:2017ffs, Monitor:2017mdv}. GW170817 has become the first standard siren detected, with redshift $z=0.008^{+0.002}_{-0.003}$ and the source was localized at luminosity distance $d_L=40^{+8}_{-14}$ Mpc. That measurement was able constraining the Hubble constant at $70.0^{+12.0}_{-8.0}~\unit{km}~\unit{s}^{-1}~\unit{Mpc}^{-1}$ \cite{Abbott:2017xzu}. Using the delay $\Delta t = 1.74\pm0.05~\unit{s}$ between the GW and the electromagnetic signal \cite{Monitor:2017mdv}, it was possible to constrain the gravitational wave speed $c_g$. Its difference relative to the light speed $c$ was found to be very close to zero, namely $-3.10^{-15}\leq~c_g/c-1~\leq~7.10^{-16}$. This result has profound implications for many modified gravity theories and dark energy models \cite{Baker:2017hug, Ezquiaga:2017ekz, Creminelli:2017sry, Sakstein:2017xjx}. For a more complete review of GW astronomy, see \cite{Ezquiaga:2018btd}.

The third generation (3G)  GW detectors are the space interferometer LISA \cite{Audley:2017drz} and ground-based interferometers, such as the Einstein Telescope (ET) \cite{Sathyaprakash:2012jk, Maggiore:2019uih} in Europe and the Cosmic Explorer in the USA \cite{Evans:2016mbw}. The ET consists of three underground detectors distributed in the form of an equilateral triangle with 10 km arms. Covering the frequency range of $1-10^4~\unit{Hz}$, it is expected to detect a rate of $10^3-10^7$ events of NS-NS and NSBH coalescence per year, but we expect to see $\mathcal{O}(10^2)$ events with electromagnetic counterpart. Forecasts using GW's which could be detected by ET was discussed by  \cite{Zhao:2010sz, Cai:2016sby} in the context of $\Lambda$CDM model. Extra parameters as cosmic opacity \cite{Qi:2019wwb}, interaction in vacuum-energy \cite{Yang:2019bpr}, interaction dark fluids \cite{Yang:2019vni}, holographic dark energy \cite{Zhang:2019ple} and modified gravitational wave propagation \cite{Belgacem:2017ihm, Belgacem:2018lbp, Belgacem:2019pkk, Belgacem:2019tbw} were also analysed in the context of gravitational wave standard siren (GW SS). In this work, we will simulate GW's in the context of a phenomenological interacting dark sector model. Our goal is to determine how GW data only will be able to constrain the model parameters and how these data can improve the constraints obtained by current cosmological probes using SnIa, BAO and CMB. In the conclusions we will compare our results with the results obtained in \cite{Yang:2019bpr, Yang:2019vni}.

The paper is organized as follows. In section II, we present the method to use GWs as standard sirens and detection with ET telescope. In section III, we present a different class of interacting models. In section IV, we explain the methodology and section V discusses our results. Finally, our conclusions are presented in section VI.

\section{Gravitational Waves as Standard Sirens}
\hspace{0.5 cm}
The GW signal can provide a measurement of the luminosity distance, thus considered a standard siren, in analogy to SnIa, which is considered a standard candle. The theoretical expression for the luminosity distance in a FLRW flat space-time is 
\begin{equation}
\label{dLz}d_L(z)=\frac{c(1+z)}{H_0}\int_0^z\frac{dz'}{E(z', \vec{\Omega})}\quad,
\end{equation}
where $E(z,\vec{\Omega})=H(z,\vec{\Omega})/H_0$ is the normalized Hubble function, which depends on the redshift $z$ and the parameter set $\vec{\Omega}$ characterizing the cosmological model. The distance modulus corresponds to a logarithmic form of luminosity distance,
\begin{equation}
\label{muz}\mu(z)=5\log_{10}\left(\frac{d_L}{1~\unit{Mpc}}\right)+25\quad.
\end{equation}

The GW amplitude depends on the so-called chirp mass of a compact binary system, defined as $\mathcal{M}_c\equiv M\eta^{3/5}$, where $M=m_1+m_2$ is the total mass of the system and $\eta=m_1m_2/M^2$ is the symmetric mass ratio. The chirp mass can be measured by GW signal phasing \cite{Zhao:2010sz, Cai:2016sby}, thus we can obtain $d_L$ from the GW amplitude. Interferometers measure the \textit{strain} $h(t)$, which is the relative difference between two distances. In transverse-traceless gauge characterized by ``plus" modes $h_+$ and ``times" modes $h_{\times}$, the strain is given by
\begin{equation}
\label{strain}h(t)=F_+(\theta,\phi,\psi)h_+(t)+F_{\times}(\theta,\phi,\psi)h_{\times}(t)\quad,
\end{equation}
where $F_{+,\times}$ are the beam pattern functions, $\psi$ is the polarization angle and $(\theta, \phi)$ are the angles of the location of the source in the sky. The ET beam pattern functions are given by
\begin{align}
F^{(1)}_+(\theta, \phi,\psi) &= \frac{\sqrt{3}}{2}\left[\frac{1}{2}(1+\cos^2\theta)\cos2\phi\cos2\psi\right.\nonumber\\
&-\left.\cos\theta\sin2\phi\sin2\psi\right]\quad,\nonumber\\
F^{(1)}_{\times}(\theta, \phi,\psi) &= \frac{\sqrt{3}}{2}\left[\frac{1}{2}(1+\cos^2\theta)\cos2\phi\sin2\psi\right.\nonumber\\
&-\left.\cos\theta\sin2\phi\cos2\psi\right]\quad.
\end{align}
Since the three interferometers are arranged in an equilateral triangle with $60^\circ$ angle with each other, the two other beam pattern functions are related to the first by $F_{+,\times}^{(2)}(\theta,\phi,\psi)=F^{(1)}_{+,\times}(\theta, \phi+2\pi/3,\psi)$ and $F_{+,\times}^{(3)}(\theta,\phi,\psi)=F^{(1)}_{+,\times}(\theta, \phi+4\pi/3,\psi)$.

It is important to make clear that from now on when we refer to chirp mass, we will be referring to the \textit{observed chirp mass}, related to the physical chirp mass by a redshift factor, i.e, $\mathcal{M}_{c,obs}=(1+z)\mathcal{M}_{c,phys}$. The Fourier transform $\mathcal{H}(f)$ of the strain $h(t)$ is
\begin{equation}
\label{Hfourier}\mathcal{H}(f)=\mathcal{A}f^{-7/6}e^{i\Psi(f)}\quad,
\end{equation}
where $\Psi(f)$ is a phase and the amplitude is given by
\begin{equation}
\label{AmpOG}\mathcal{A}=\frac{1}{d_L}\sqrt{F^2_+(1+\cos\iota)^2+4F^2_{\times}\cos\iota}\times\sqrt{\frac{5\pi}{96}}\pi^{-7/6}\mathcal{M}_c^{5/6}\quad,
\end{equation}
where $\iota$ is the angle between the angular orbital momentum and the line of sight.

We will generate a mock catalog $d_L-z$ by coalescence of BNS pair in the mass range $[1-2]M_{\odot}$ for each individual neutron star. The redshift distribution of the observable sources follow the function \cite{Belgacem:2019tbw}
\begin{equation}
\label{Pzdist}P(z)=\frac{R_z(z)}{\int_0^{10} R_z(z)~dz}\quad,
\end{equation}
where  $R_z(z)$ describes the redshift evolution of burst rate per unity of redshift. It takes the form
\begin{equation}
\label{RatioZ}R_z(z)=\frac{R_{\unit{BNS}}(z)}{1+z}\frac{dV(z)}{dz}\quad,
\end{equation}
where $dV/dz$ is the comoving volume element and $R_{\unit{BNS}}(z)$ is the rate per volume in the source frame, which can be modeled by combining the formation rate of massive binaries $R_F$ and the delay time distribution $P(t_d)$ \cite{Belgacem:2019tbw, Vangioni:2015ofa, Howell:2018nhu, Vitale:2018yhm}:
\begin{equation}
    \label{ratioBNS}R_{\unit{BNS}}(z)=\int_{t_{\unit{min}}}^{t_{\unit{max}}}R_F(z_f)P(t_d)dt_d\quad.
\end{equation}
$P(t_d)\propto t_d^{-1}$ for $t_d>t_{\unit{min}}$, $t_{\unit{min}}=20~\unit{Myr}$ is the minimal delay time for a BNS system to evolve to merger, $t_{\unit{max}}$ is the Hubble time and $z_f$ is the binary pair formation redshift. For $R_{\unit{BNS}}(z)$ was adopted the cosmic star formation rate based in Gamma-Ray Burst (GRB) rate from \cite{Vangioni:2014axa}. For the present time rate we adopted $R_{\unit{BNS}}(z=0)=920~\unit{Gpc}^{-3}\unit{yr}^{-1}$ estimated by O1/O2 LIGO/Virgo observation run with the assumption that the mass distribution of neutron stars follow a gaussian mass distribution \cite{Abbott:2017gyy}.
Following \cite{Zhao:2010sz,Cai:2016sby}, since the maximal inclination is $\iota=20^\circ$, we consider $\iota= 0^{\circ}$ and assume that the amplitude given by eq. \eqref{AmpOG} does not depend on the polarization angle $\psi$.

To perform the complete simulation, we need the noise power spectral density $S_h(f)$ (PSD) of ET given in \cite{Zhao:2010sz} to calculate the Signal-to-Noise ratio (SNR) of the network of three independent interferometers
\begin{equation}
\label{rhoSNR}\rho=\sqrt{\sum_{i=1}^3(\rho^{(i)})^2}\quad,
\end{equation}
where $\rho^{(i)}=\sqrt{\langle\mathcal{H}^{(i)},\mathcal{H}^{(i)}\rangle}$. The inner product of two functions $a(t)$ and $b(t)$ is defined as
\begin{equation}
\label{innerprod}\langle a,b \rangle = 4\int_{f_{\unit{lower}}}^{f_{\unit{upper}}}\frac{\tilde{a}(f)\tilde{b}^*(f)+\tilde{a}^*(f)\tilde{b}(f)}{2}\frac{df}{S_h(f)}\quad,
\end{equation}
where $\tilde{a}(f)$ and $\tilde{b}(f)$ are, respectively, the Fourier transforms of $a(t)$ and $b(t)$. The lower limit in frequency of ET is $f_{\unit{lower}}=1$ Hz and the upper limit is given by $f_{\unit{upper}}=2/(6^{3/2}2\pi M_{\unit{obs}})$ where $M_{\unit{obs}}=(1+z)M_{\unit{phys}}$ is the observed total mass \cite{Zhao:2010sz}.

The standard Fisher matrix method is used to estimate the instrumental error in luminosity distance, assuming that this parameter is uncorrelated with any other GW parameters \cite{Li:2013lza}, such that
\begin{equation}
\label{fisherinst}\sigma_{d_L}^{\unit{inst}}\simeq\sqrt{\left\langle\frac{\partial\mathcal{H}}{\partial d_L},\frac{\partial\mathcal{H}}{\partial d_L}\right\rangle^{-1}}\quad.
\end{equation}
Since $\mathcal{H}\propto d_L^{-1}$, we have $\sigma_{d_L}^{\unit{inst}}\simeq d_L/\rho$. To take into account the effect of inclination $\iota$, where $0^\circ <\iota<90^\circ$, we add a factor of 2 in the instrumental error. Therefore,
\begin{equation}
\label{sigmainst}\sigma_{d_L}^{\unit{inst}}\simeq\frac{2d_L}{\rho}\quad.
\end{equation}
We have to consider an additional error due to gravitational lensing. For ET, this error is $\sigma_{d_L}^{\unit{lens}}=0.05~zd_L$. Thus, the total uncertainty on luminosity distance is
\begin{align}
\label{sigmadLtot}\sigma_{d_L} &= \sqrt{(\sigma_{d_L}^{\unit{inst}})^2+(\sigma_{d_L}^{\unit{lens}})^2}\nonumber\\
&= \sqrt{\left(\frac{2d_L}{\rho}\right)^2+(0.05~zd_L)^2}\quad.
\end{align}
The uncertainty in the distance modulus $\eqref{muz}$ is propagated from the uncertainty in luminosity distance $\eqref{sigmadLtot}$ as
\begin{equation}
\label{sigmamu}\sigma_{\mu}=\frac{5}{\ln10}\frac{\sigma_{d_L}}{d_L}\quad,
\end{equation}
which we use to generate the mock error bars in the distance modulus catalog.

\section{Interacting Dark Sector Scenario}
\hspace{0.5 cm}
We consider a homogeneous and isotropic background described by a spacial flat Friedmann-Lema\^{i}tre-Robertson-Walker (FLRW) metric. The total energy  density $\rho_{tot}$ consists of four species: $\rho_{tot}=\rho_{dm}+\rho_{de}+\rho_b+\rho_r$ 
where ``$dm$" denotes dark matter, ``$de$" denotes dark energy, ``$b$"  baryons and ``$r$"  radiation (photons and neutrinos). Since the nature of DM and DE are still unknown and they dominate the energy content of the universe today, it is reasonable to consider that the two components of the dark sector can interact with each other. However, the coupling must be small in view of the fact that the $\Lambda$CDM model agrees very well with the data and the interacting model can not deviate much from the $\Lambda$CDM predictions.

In this model, baryons and radiation evolve independently of the other components, but dark matter and dark energy evolve following the coupled conservation equations
\begin{align}
\label{dmcontinuity}\dot{\rho}_{dm}+3H\rho_{dm} &= Q\quad,\\
\label{decontinuity}\dot{\rho}_{de}+3H(1+\omega)\rho_{de} &= -Q\quad,
\end{align} 
in such a way that the total energy density of the dark sector is still conserved. In equations \eqref{dmcontinuity} and \eqref{decontinuity}, a dot represents derivative with respect to the cosmic time, $\omega$ represents the dark energy equation of state and $Q$ is the coupling. Note that $Q>0$ means that the energy transfers from dark energy to dark matter and for $Q<0$ we have the opposite. By dimensional analysis, we know that the coupling function $Q$ must have dimension of energy density $\rho$ over time $t$. We consider three phenomenological models: Model I, where $Q=3H\xi\rho_{dm}$; Model II, where $Q=3H\xi\rho_{de}$; and Model III where $Q=3H\xi(\rho_{dm}+\rho_{de})$. Here, $\xi$ is the coupling constant.

\subsection{Model I}
\hspace{0.5 cm}
\label{SecModI}
For this model, we can solve the system of equations given by eq. \eqref{dmcontinuity} and eq. \eqref{decontinuity} and obtain the analytical solution as a function of redshift $z$ for dark matter and dark energy densities, respectively,
\begin{align}
\label{rhodmM1}\rho_{dm}(z) &= \rho_{dm,0}(1+z)^{3(1-\xi)}\quad,\\
\label{rhodeM1}\rho_{de}(z) &= \left(\rho_{de,0}+\frac{\xi}{\xi+\omega}\rho_{dm,0}\right)(1+z)^{3(1+\omega)}-\frac{\xi}{\xi+\omega}\rho_{dm,0}(1+z)^{3(1-\xi)}\quad.
\end{align}
The normalized Hubble function $E(z)=H(z)/H_0$ is given by the expression
\begin{align}
\label{HzModel1}E(z)^2 &= \Omega_{b,0}(1+z)^3 +\Omega_{r,0}(1+z)^4\nonumber\\
&+\frac{\omega}{\xi+\omega}\Omega_{dm,0}(1+z)^{3(1-\xi)}+\left(\Omega_{de,0}+\frac{\xi}{\xi+\omega}\Omega_{dm,0}\right)(1+z)^{3(1+\omega)}\quad.
\end{align}

\subsection{Model II}
\hspace{0.5 cm}
Here, the evolution of dark energy and dark matter densities are, respectively,
\begin{align}
\label{rhodeM2}\rho_{de}(z) &= \rho_{de,0}(1+z)^{3(1+\xi+\omega)}\quad,\\
\label{rhodmM2}\rho_{dm}(z) &= \left(\rho_{dm,0}+\frac{\xi}{\xi+\omega}\rho_{de,0}\right)(1+z)^3-\frac{\xi}{\xi+\omega}\rho_{de,0}(1+z)^{3(1+\xi+\omega)}\quad,
\end{align}
and the normalized Hubble function is given by the expression
\begin{align}
\label{HzModel2}E(z)^2 &= \Omega_{b,0}(1+z)^3 +\Omega_{r,0}(1+z)^4\nonumber\\
&+\left(\Omega_{dm,0}+\frac{\xi}{\xi+\omega}\Omega_{de,0}\right)(1+z)^3+\frac{\omega}{\xi+\omega}\Omega_{de,0}(1+z)^{3(1+\xi+\omega)}\quad.
\end{align}

\subsection{Model III}
\hspace{0.5 cm}
For this model, we solve \eqref{dmcontinuity} and \eqref{decontinuity} numerically with a modified version of the CAMB code  \cite{Lewis:1999bs} to obtain dark matter and dark energy densities, Hubble function and luminosity distance.

\section{Methodology}
\hspace{0.5 cm}
As prescript by \cite{Belgacem:2019tbw}, we generate a ``realistic" and an ``optimistic" joint GW-GRB sampling, with 60 and 600 events, respectively. The distribution of events follows eq. $\eqref{Pzdist}$ in a redshift range $(0,2)$ and we calculate the respective $d_L(z)$ and $\mu(z)$ of each event assuming each interacting model as a fiducial model.  We consider as fiducial values the best-fit parameters obtained with Planck2015 + BAO + SNIa + $H_0$ data in reference \citep{Costa:2016tpb} (see Table \ref{FidValue}). The Model II will be split in two parts: Model IIA with $\omega<-1$ and Model IIB with $-1<\omega<-1/3$. Model I and Model III are restricted to $\omega<-1$. These constraints are due to instability in curvature perturbations \cite{He:2008si, Gavela:2009cy}. 
\begin{table}
\begin{center}
\label{TableI}
\begin{tabular}{|c|c|c|c|c|}
 \hline
 Parameter & Model I & Model IIA & Model IIB & Model III\\
 \hline
 $\Omega_m$ & 0.312 & 0.3265 & 0.2351 & 0.3149 \\
 \hline
 $H_0$ & 67.93 & 68.76 & 68.45 & 67.59 \\
 \hline
 $\omega$ & -1.06 & -1.087 & -0.9434 & −1.051\\
 \hline
 $\xi$ & 0.0007273 & 0.03798 & -0.09291 & 0.001205\\
 \hline
\end{tabular}
\end{center}
\caption{Fiducial values}
\label{FidValue}
\end{table}

\begin{figure}[h]
\centering
\includegraphics[totalheight=1.7in]{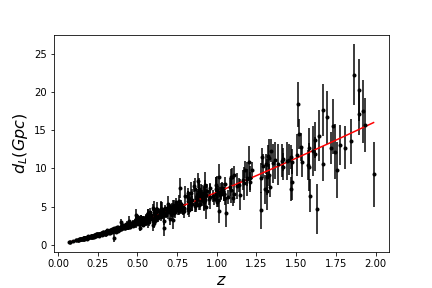}
\caption{Simulated $d_L(z) - z$ catalog for ``optimistic" GW-GRB joint detections. The red line shows the fiducial luminosity distance for IM1}\label{dLIM1}
\qquad
\includegraphics[totalheight=1.7in]{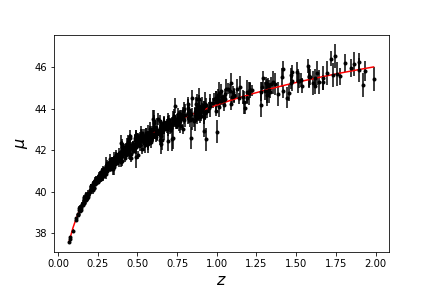}
\caption{Simulated $\mu(z) - z$ catalog for ``optimistic" GW-GRB joint detections. The red line shows the fiducial distance modulus for IM1}\label{muIM1}
\end{figure}

We randomly generate the mass of neutron stars in the interval $[1-2]M_{\odot}$. The position angles $\theta$ and $\phi$ are in the intervals  $[0-\pi]$ and $[0-2\pi]$, respectively. Then, we calculate the SNR for the three detectors given by eq. $\eqref{rhoSNR}$ for each set of random sample and confirm the detection if $\rho_{net}>8.0$. If the detection is confirmed, we calculate the errors $\sigma_{d_L}$ and $\sigma_{\mu}$ by eq. $\eqref{sigmadLtot}$ and eq. $\eqref{sigmamu}$, respectively. Finally, we consider as the ``real" detection a Gaussian dispersion around the fiducial values, i.e, $d_L^{\unit{real}}=\mathcal{N}(d_L^{\unit{fid}},\sigma_{d_L})$ and $\mu^{\unit{real}}=\mathcal{N}(\mu^{\unit{fid}},\sigma_{\mu})$. Thus, we can simulate a sampling of GW sources  with their respective luminosity distance and redshift as we can see in Fig.~\ref{dLIM1}, where we simulate an ``optimistic" catalog. In Figure~\ref{muIM1}, we show the $\mu(z)-z$ simulated catalog for the same catalog.

After generating the sampling, we are able to constrain the set of model parameters $\vec{\Omega}=\{\Omega_m, H_0, \omega, \xi\}$. We calculate the $\chi^2$ for $N$ simulated data points, given by
\begin{equation}
\label{chi2}\chi^2=\sum_{i=1}^N\left[\frac{\bar{\mu}^i-\mu(\bar{z}_i;\vec{\Omega})}{\bar{\sigma}_{\mu}^i}\right]^2\quad,
\end{equation} 
where $\bar{z}_i$, $\bar{\mu}^i$ and $\bar{\sigma}_{\mu}^i$ are, respectively, the $i$th redshit, distance modulus and error on the distance modulus of each simulated data set. 

\section{Results}
\hspace{0.5 cm}
 In this section, we present the constraints obtained for each interacting model and compare GW with current cosmological probes. Therefore, we calculate the constraints using the latest data from the Planck satellite mission \cite{Aghanim:2018eyx}. We consider Planck 2018 measurements from high-$\ell$ multipole temperature and polarization data, TT + TE + EE, and also low-$\ell$ temperature only \textrm{Commander} likelihood plus EE \textrm{SimAll} likelihood. We also combine the CMB measurements from Planck with 1048 SnIa data from the latest Pantheon sample \cite{Scolnic:2017caz} and five Baryon Acoustic Oscillations (BAO) data. We use the 6dFGS and SDSS-MGS measurements at effective redshifts $z_{\unit{eff}} = 0.106$ and $z_{\unit{eff}} = 0.15$, respectively  \cite{Beutler:2011hx, Ross:2014qpa}, and the latest BOSS data release 12 summarized in \cite{Alam:2016hwk} at redshifts $z_{\unit{eff}} = 0.38$, $0.51$ and $0.61$. We ran the MCMC algorithm with CMB \textit{Planck} 2018 data only, CMB+BAO and CMB+SN. Then, we add the GW ``realistic" and ``optmistic" simulated data to CMB and to CMB+BAO+SN in order to determine how GW data can improve the constraints obtained with current cosmological probes. The analyse using CMB data were made with modified versions of the CAMB code \cite{Lewis:1999bs} and the CosmoMC code \cite{Lewis:2002ah,Lewis:2013hha}. We set the statistical convergence according to the Gelman and Rubin criterion as $R - 1 = 0.03$ \cite{Gelman:1992zz}.

\subsection{Model I}
\hspace{0.5 cm}
In Table~\ref{ModelIDatasetsGWRl}, we list the average and $68\%$ confidence levels (C.L.) for the parameters of Model I (Sec.\ref{SecModI}). There we compare the constraints for different datasets with our ``realistic" catalog, while in Table~\ref{ModelIDatasetsOpt} we compare the constraints using the ``realistic" or ``optimistic" simulated data. For the coupling constant $\xi$, CMB+GW Real. provide an improvement to CMB data only ($\sim 27\%$) which is less than CMB+BAO ($\sim 55\%$), but superior to CMB+SN ($\sim 9\%$). For $H_0$, GW presents a very restrictive power, with improvements in relation to CMB data of $\sim 88\%$, which is better than CMB+BAO  ($\sim 85\%$) and CMB+SN ($\sim 86\%$). With respect to $\Omega_m$, CMB+GW Real. have a more restrictive power ($\sim 79\%)$ than CMB+BAO ($\sim 75\%)$ and CMB+SN ($\sim 54\%$), while for the EoS $\omega$ it is inferior: $\sim 71\%$ in comparison to CMB+BAO ($\sim 81\%$) and SN  ($\sim 82\%$). Combining all data together, we obtain an improvement in the parameter constraints with respect to Planck only of: ($\sim 66\%$) for the interaction $\xi$, ($\sim 94\% $) for the Hubble constant, ($\sim 88\%$) for the matter fraction $\Omega_m$ and ($\sim 92\%$) for the dark energy EoS.

\begin{figure}[!ht]
\centering
\begin{center}
\hspace{3.0cm}
\includegraphics[totalheight=3.5in]{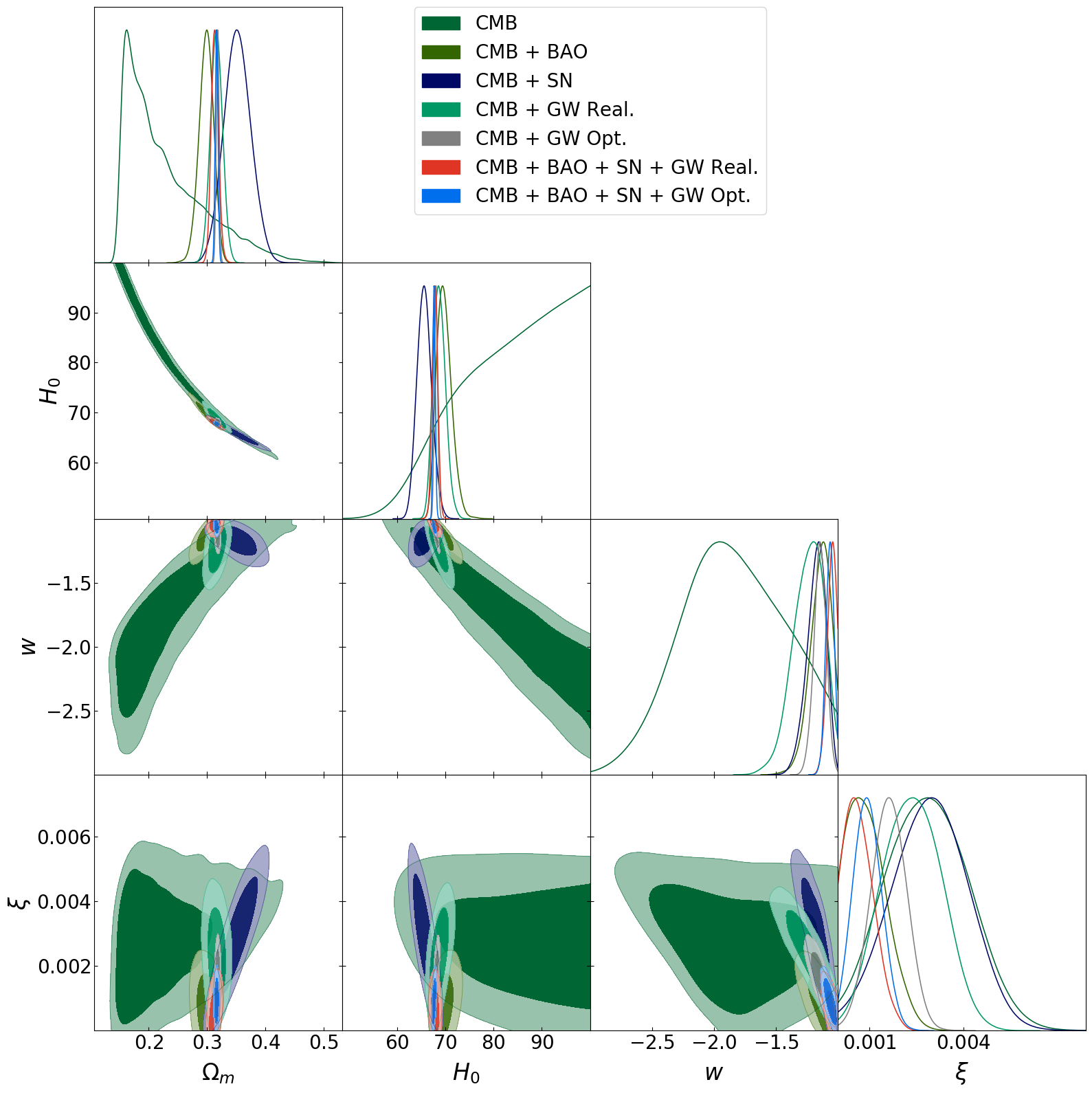}
\caption{\footnotesize 1D and 2D confidence contours for IM1.}
\label{contoursIM1GW}
\end{center}
\end{figure}

When we consider CMB+GW Opt., we obtain an improvement in the coupling of $\sim 42\%$ with respect to CMB+GW Real., but no significant improvement between CMB+BAO+SN+GW Opt. and CMB+BAO+SN+GW Real. Meanwhile, for the Hubble constant, the addition of an optimistic catalog for CMB and CMB+BAO+SN restricts in $\sim 44\%$ and $\sim 57\%$ with respect to the addition of GW Real. to the same datasets, respectively. For $\Omega_m$, they are respectively $\sim 76\%$ and $\sim 61\%$, and for $\omega$ they are $\sim 59\%$ and $\sim 3\%$. Figure~\ref{contoursIM1GW} presents the 1-D and 2-D confidence level contours for those parameters using different data configurations.

\begin{table}
\footnotesize
\begin{center}
\begin{tabular}{|c|r|r|r|r|r|}
\cline{2-6}
\multicolumn{1}{c}{}  & \multicolumn{1}{|c|}{\footnotesize{ CMB}} & \multicolumn{1}{|c|}{\footnotesize{CMB + BAO}} & \multicolumn{1}{|c|}{\footnotesize{CMB + SN}} & \multicolumn{1}{|c|}{\scriptsize{CMB + GW Rl}.} & 
\multicolumn{1}{|c|}{\tiny{CMB+BAO+SN+GW Rl.}} \\ 
 \hline
 
Parameter & Avg. $\pm$ 68\% limits & Avg. $\pm$ 68\% limits & Avg. $\pm$ 68\% limits  & Avg. $\pm$ 68\% limits & Avg. $\pm$ 68\% limits \\  
 \hline
$\Omega_b h^2$ & $0.02262^{+0.000183}_{-0.000181}$ & $0.02247^{+0.000157}_{-0.000156}$ & $0.02257^{+0.00018}_{-0.000177}$ & $0.02256^{+0.000173}_{-0.000178}$ & 
$0.02245^{+0.000152}_{-0.00015}$ \\ 
$\Omega_c h^2$ & $0.1273^{+0.00366}_{-0.00375}$ & $0.1221^{+0.00157}_{-0.00174}$ & $0.1282^{+0.00343}_{-0.00346}$ & $0.1263^{+0.00269}_{-0.00247}$ & 
$0.1213^{+0.00119}_{-0.00138}$ \\ 
$100\theta_{MC}$ & $1.04^{+0.000379}_{-0.000375}$ & $1.041^{+0.000297}_{-0.000294}$ & $1.04^{+0.000374}_{-0.000366}$ & $1.041^{+0.000334}_{-0.000332}$ & 
$1.041^{+0.000288}_{-0.000291}$ \\ 
$\tau$ & $0.05154^{+0.00711}_{-0.00717}$ & $0.05455^{+0.00692}_{-0.00783}$ & $0.05181^{+0.00761}_{-0.00752}$ & $0.05204^{+0.00735}_{-0.00747}$ & 
$0.05494^{+0.00759}_{-0.00747}$ \\ 
${\rm{ln}}(10^{10}A_s)$ & $3.034^{+0.0157}_{-0.0153}$ & $3.042^{+0.0153}_{-0.0155}$ & $3.035^{+0.0164}_{-0.0165}$ & $3.036^{+0.0156}_{-0.016}$ & 
$3.043^{+0.0159}_{-0.0157}$ \\ 
$n_s$ & $0.9598^{+0.00492}_{-0.00491}$ & $0.9642^{+0.00399}_{-0.00399}$ & $0.9593^{+0.00463}_{-0.00468}$ & $0.9608^{+0.0043}_{-0.00437}$ & 
$0.9653^{+0.00405}_{-0.00401}$ \\ 
$w$ & $-1.842^{+0.426}_{-0.439}$ & $-1.145^{+0.103}_{-0.0607}$ & $-1.176^{+0.0891}_{-0.0623}$ & $-1.243^{+0.146}_{-0.101}$ & 
$-1.056^{+0.0475}_{-0.0202}$ \\ 
$\xi$ & $0.002881^{+0.00127}_{-0.0014}$ & $0.0009622^{+0.000355}_{-0.000842}$ & $0.002942^{+0.00119}_{-0.00123}$ & $0.002368^{+0.000963}_{-0.000974}$ & 
$0.0007218^{+0.000257}_{-0.000637}$ \\ 
\hline
$H_0$ & 
$83.72^{+16.3}_{-5.4}$ & 
$69.8^{+1.28}_{-1.77}$ & 
$65.68^{+1.39}_{-1.58}$ & 
$68.8^{+1.17}_{-1.38}$ & 
$67.88^{+0.574}_{-0.578}$ \\
$\Omega_{de}$ & $0.7727^{+0.0791}_{-0.0227}$ & $0.7016^{+0.0115}_{-0.013}$ & $0.6481^{+0.0231}_{-0.0228}$ & $0.6839^{+0.0102}_{-0.0103}$ & 
$0.6866^{+0.00642}_{-0.00575}$ \\ 
$\Omega_m$ & $0.2273^{+0.0227}_{-0.0791}$ & $0.2984^{+0.013}_{-0.0115}$ & $0.3519^{+0.0228}_{-0.0231}$ & $0.3161^{+0.0103}_{-0.0102}$ & 
$0.3134^{+0.00575}_{-0.00642}$ \\ 
$\sigma_8$ & $0.8499^{+0.331}_{-0.863}$ & $0.8348^{+0.0215}_{-0.0179}$ & $0.7787^{+0.21}_{-0.83}$ & $0.7651^{+0.108}_{-0.717}$ & 
$0.8173^{+0.0108}_{-0.0121}$ \\ 
${\rm{Age}}/{\rm{Gyr}}$ & $13.83^{+0.149}_{-0.193}$ & $13.81^{+0.0377}_{-0.0451}$ & $14.03^{+0.109}_{-0.111}$ & $13.94^{+0.0615}_{-0.0622}$ & 
$13.83^{+0.0277}_{-0.0374}$ \\ 
 \hline
\end{tabular}
\end{center}
\caption{Model I - Cosmological Parameters for different datasets + GW Real.}
\label{ModelIDatasetsGWRl}
\end{table}

\begin{table}
\footnotesize
\begin{center}
\begin{tabular}{|c|r|r|r|r|r|}
\cline{2-5}
\multicolumn{1}{c}{}  & 
\multicolumn{1}{|c|}{\scriptsize{CMB + GW Rl}.} & 
\multicolumn{1}{|c|}{\scriptsize{CMB + GW Opt.}} & 
\multicolumn{1}{|c|}{\tiny{CMB+BAO+SN+GW Rl.}} & 
\multicolumn{1}{|c|}{\tiny{CMB+BAO+SN+GW Opt.}} \\
 \hline
 
Parameter & Avg. $\pm$ 68\% limits & Avg. $\pm$ 68\% limits & Avg. $\pm$ 68\% limits  & Avg. $\pm$ 68\% limits \\  
 \hline
$\Omega_b h^2$ & 
$0.02256^{+0.000173}_{-0.000178}$ & 
$0.02249^{+0.000174}_{-0.000168}$ &
$0.02245^{+0.000152}_{-0.00015}$ & 
$0.02245^{+0.000159}_{-0.00016}$ \\
$\Omega_c h^2$ & 
$0.1263^{+0.00269}_{-0.00247}$ & 
$0.1244^{+0.00147}_{-0.00148}$ &
$0.1213^{+0.00119}_{-0.00138}$ & 
$0.1219^{+0.00114}_{-0.00117}$ \\
$100\theta_{MC}$ & 
$1.041^{+0.000334}_{-0.000332}$ & 
$1.041^{+0.000313}_{-0.000302}$ & 
$1.041^{+0.000288}_{-0.000291}$ & 
$1.041^{+0.000293}_{-0.000289}$\\
$\tau$ & 
$0.05204^{+0.00735}_{-0.00747}$ & 
$0.05288^{+0.00695}_{-0.00751}$ &
$0.05494^{+0.00759}_{-0.00747}$ & 
$0.0549^{+0.00724}_{-0.00784}$ \\
${\rm{ln}}(10^{10}A_s)$ & 
$3.036^{+0.0156}_{-0.016}$ & 
$3.038^{+0.0145}_{-0.0156}$ &
$3.043^{+0.0159}_{-0.0157}$ & 
$3.042^{+0.0159}_{-0.0159}$ \\
$n_s$ & 
$0.9608^{+0.0043}_{-0.00437}$ & 
$0.9618^{+0.0041}_{-0.00413}$ & 
$0.9653^{+0.00405}_{-0.00401}$ &
$0.9649^{+0.00393}_{-0.00389}$\\
$w$ & 
$-1.243^{+0.146}_{-0.101}$ & 
$-1.146^{+0.053}_{-0.0465}$ & 
$-1.056^{+0.0475}_{-0.0202}$ & 
$-1.068^{+0.036}_{-0.0293}$\\
$\xi$ & 
$0.002368^{+0.000963}_{-0.000974}$ & 
$0.001622^{+0.00056}_{-0.000562}$ & 
$0.0007218^{+0.000257}_{-0.000637}$ & 
$0.0009523^{+0.000415}_{-0.000487}$\\
\hline
$H_0$ & 
$68.8^{+1.17}_{-1.38}$ & 
$68.09^{+0.321}_{-0.323}$ & 
$67.88^{+0.574}_{-0.578}$ & 
$67.71^{+0.227}_{-0.26}$\\
$\Omega_{de}$ & 
$0.6839^{+0.0102}_{-0.0103}$ & 
$0.6818^{+0.00237}_{-0.00235}$ & 
$0.6866^{+0.00642}_{-0.00575}$ & 
$0.6837^{+0.00234}_{-0.00232}$\\
$\Omega_m$ & 
$0.3161^{+0.0103}_{-0.0102}$ & 
$0.3182^{+0.00235}_{-0.00237}$ & 
$0.3134^{+0.00575}_{-0.00642}$ & 
$0.3163^{+0.00232}_{-0.00234}$\\
$\sigma_8$ & 
$0.7651^{+0.108}_{-0.717}$ & 
$0.8219^{+0.0222}_{-0.00372}$ & 
$0.8173^{+0.0108}_{-0.0121}$ & 
$0.8179^{+0.0107}_{-0.0101}$\\
${\rm{Age}}/{\rm{Gyr}}$ & 
$13.94^{+0.0615}_{-0.0622}$ & 
$13.89^{+0.0325}_{-0.0318}$ & 
$13.83^{+0.0277}_{-0.0374}$ & 
$13.85^{+0.0232}_{-0.0269}$\\
 \hline
\end{tabular}
\end{center}
\caption{Model I - Cosmological Parameters for GW Real. and GW Opt.}
\label{ModelIDatasetsOpt}
\end{table}

\subsection{Model IIA}
\begin{figure}[!ht]
\centering
\begin{center}
\hspace{3.0cm}
\includegraphics[totalheight=3.5in]{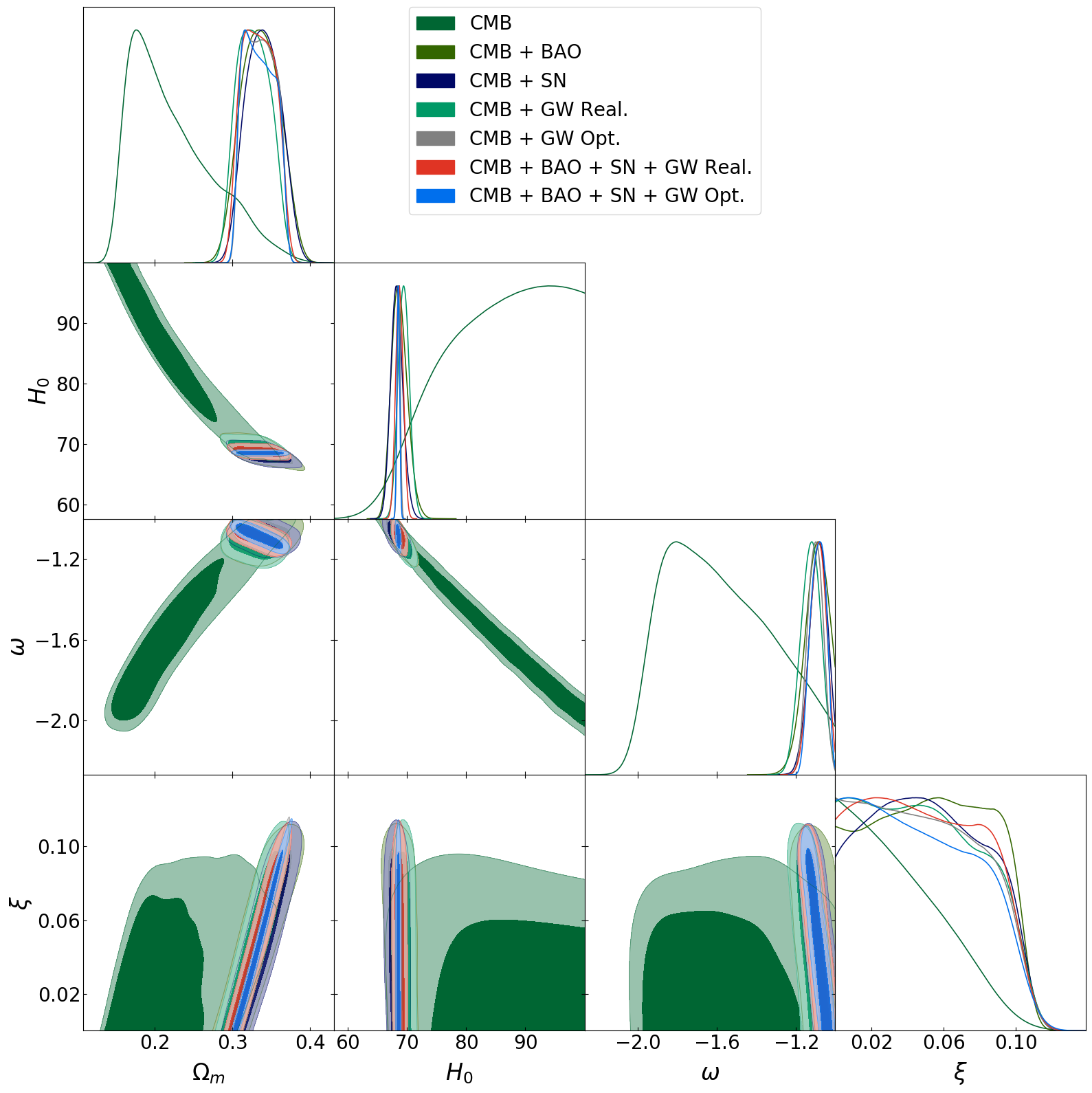}
\caption{\footnotesize 1D and 2D confidence contours for IM2A.}
\label{contoursIM2AGW}
\end{center}
\end{figure}

\hspace{0.5 cm}
Again we consider the effect of GW measurements with a ``realistic" and compare with an "optimistic" catalog, which are shown in Table~\ref{ModelIIADatasetsGWRl} and Table~\ref{ModelIIADatasetsGWOpt}, respectively. Figure~\ref{contoursIM2AGW} shows 1-D and 2-D confidence contours. We can see that the addition of GW's do not restrict significantly the contours in $\xi$ in relation to other datasets. However, for the other parameters, the addition of GW's has a remarkable effect. In special for $H_0$, we have improvements with relation to CMB data of $\sim 64\%$ to CMB+GW Real., which is better than CMB+BAO  ($\sim 58\%$) and worse than CMB+SN ($\sim 89\%$). With respect to $\Omega_m$, CMB+GW Real. have a more restrictive power ($\sim 54\%)$ than CMB+BAO ($\sim 47\%)$ and CMB+SN ($\sim 51\%$), while for the EoS $\omega$ is $\sim 81\%$ in comparison to CMB+BAO ($\sim 78\%$) and CMB+SN ($\sim 83\%$). The combination of those data improve our constraints with respect to Planck only as:  ($\sim 93\%$) for $H_0$, ($\sim 54\%$) for the matter fraction and ($\sim 84\%$) for the dark energy EoS.

There is no significant improvement to CMB+GW Opt. in relation to CMB+GW Real. for the coupling parameter, but there is an improvement of $\sim 4\%$ to CMB+BAO+SN+GW Opt. in relation to CMB+BAO+SN+GW Real., while for $H_0$ the restriction is $\sim 73\%$ and $\sim 63\%$ better, respectively. For $\Omega_m$, they are respectively $\sim 6\%$ and $\sim 4\%$, and for $\omega$ are respectively $\sim 22\%$ and $\sim 6\%$.

\begin{table}
\footnotesize
\begin{center}
\begin{tabular}{|c|r|r|r|r|r|}
\cline{2-6}
\multicolumn{1}{c}{}  & \multicolumn{1}{|c|}{\footnotesize{ CMB}} & \multicolumn{1}{|c|}{\footnotesize{CMB + BAO}} & \multicolumn{1}{|c|}{\footnotesize{CMB + SN}} & \multicolumn{1}{|c|}{\scriptsize{CMB + GW Rl}.} & 
\multicolumn{1}{|c|}{\tiny{CMB+BAO+SN+GW Rl.}} \\ 
 \hline
 
Parameter & Avg. $\pm$ 68\% limits & Avg. $\pm$ 68\% limits & Avg. $\pm$ 68\% limits  & Avg. $\pm$ 68\% limits & Avg. $\pm$ 68\% limits \\  
 \hline
$\Omega_b h^2$ & $0.02238^{+0.000153}_{-0.000152}$ & $0.02238^{+0.000143}_{-0.000145}$ & $0.02236^{+0.000149}_{-0.000149}$ & $0.02235^{+0.000141}_{-0.000143}$ & 
$0.02242^{+0.000141}_{-0.000141}$ \\ 
$\Omega_c h^2$ & $0.1333^{+0.00614}_{-0.0121}$ & $0.1357^{+0.0115}_{-0.00745}$ & $0.1354^{+0.00814}_{-0.0105}$ & $0.135^{+0.00761}_{-0.0117}$ & 
$0.1347^{+0.0105}_{-0.0115}$ \\ 
$100\theta_{MC}$ & $1.04^{+0.000628}_{-0.000525}$ & $1.04^{+0.000567}_{-0.000556}$ & $1.04^{+0.000556}_{-0.000552}$ & $1.04^{+0.000578}_{-0.000573}$ & 
$1.04^{+0.000585}_{-0.000588}$ \\ 
$\tau$ & $0.05429^{+0.00719}_{-0.0078}$ & $0.05488^{+0.00759}_{-0.00762}$ & $0.05427^{+0.00735}_{-0.00793}$ & $0.05395^{+0.0073}_{-0.0074}$ & 
$0.05535^{+0.00762}_{-0.00758}$ \\ 
${\rm{ln}}(10^{10}A_s)$ & $3.044^{+0.0148}_{-0.016}$ & $3.045^{+0.0159}_{-0.0157}$ & $3.045^{+0.0154}_{-0.0155}$ & $3.044^{+0.0146}_{-0.0162}$ & 
$3.045^{+0.0156}_{-0.0156}$ \\ 
$n_s$ & $0.965^{+0.00429}_{-0.00426}$ & $0.9651^{+0.0042}_{-0.00422}$ & $0.9643^{+0.00431}_{-0.00432}$ & $0.9642^{+0.00416}_{-0.00421}$ & 
$0.9658^{+0.00386}_{-0.00422}$ \\ 
$w$ & $-1.586^{+0.197}_{-0.345}$ & $-1.099^{+0.0705}_{-0.0439}$ & $-1.088^{+0.0496}_{-0.0399}$ & $-1.125^{+0.0536}_{-0.0481}$ & 
$-1.089^{+0.0449}_{-0.0381}$ \\ 
$\xi$ & $0.03718^{+0.0112}_{-0.0372}$ & $0.05359^{+0.04}_{-0.029}$ & $0.0517^{+0.0233}_{-0.0434}$ & $0.04985^{+0.0163}_{-0.0498}$ & 
$0.05077^{+0.0177}_{-0.0508}$ \\ 
\hline
$H_0$ & $85.36^{+14.6}_{-4.73}$ & $68.75^{+1.18}_{-1.48}$ & $68.3^{+0.929}_{-1.09}$ & $69.47^{+0.93}_{-0.947}$ & 
$68.74^{+0.605}_{-0.647}$ \\ 
$\Omega_{de}$ & $0.7775^{+0.0676}_{-0.0271}$ & $0.6636^{+0.0259}_{-0.024}$ & $0.6602^{+0.0229}_{-0.0229}$ & $0.6723^{+0.0235}_{-0.0199}$ & 
$0.6661^{+0.0229}_{-0.0199}$ \\ 
$\Omega_m$ & $0.2225^{+0.0271}_{-0.0676}$ & $0.3364^{+0.024}_{-0.0259}$ & $0.3398^{+0.0229}_{-0.0229}$ & $0.3277^{+0.0199}_{-0.0235}$ & 
$0.3339^{+0.0199}_{-0.0229}$ \\ 
$\sigma_8$ & $0.8924^{+0.0874}_{-0.0808}$ & $0.7587^{+0.0359}_{-0.0476}$ & $0.76^{+0.0333}_{-0.0425}$ & $0.7703^{+0.0381}_{-0.0429}$ & 
$0.7589^{+0.0346}_{-0.0437}$ \\ 
${\rm{Age}}/{\rm{Gyr}}$ & $13.56^{+0.0575}_{-0.129}$ & $13.77^{+0.029}_{-0.0294}$ & $13.78^{+0.0275}_{-0.0272}$ &  $13.76^{+0.0217}_{-0.0214}$ & 
$13.77^{+0.0187}_{-0.0188}$ \\ 
 \hline
\end{tabular}
\end{center}
\caption{Model IIA - Cosmological Parameters for different datasets + GW Real.}
\label{ModelIIADatasetsGWRl}
\end{table}

\begin{table}
\footnotesize
\begin{center}
\begin{tabular}{|c|r|r|r|r|r|}
\cline{2-5}
\multicolumn{1}{c}{}  & 
\multicolumn{1}{|c|}{\scriptsize{CMB + GW Rl}.} & 
\multicolumn{1}{|c|}{\scriptsize{CMB + GW Opt.}} & 
\multicolumn{1}{|c|}{\tiny{CMB+BAO+SN+GW Rl.}} & 
\multicolumn{1}{|c|}{\tiny{CMB+BAO+SN+GW Opt.}} \\
 \hline
 
Parameter & Avg. $\pm$ 68\% limits & Avg. $\pm$ 68\% limits & Avg. $\pm$ 68\% limits  & Avg. $\pm$ 68\% limits \\  
 \hline
$\Omega_b h^2$ & 
$0.02235^{+0.000141}_{-0.000143}$ & 
$0.02232^{+0.000137}_{-0.000139}$ & $0.02242^{+0.000141}_{-0.000141}$ & 
$0.02236^{+0.00013}_{-0.000134}$\\
$\Omega_c h^2$ & 
$0.135^{+0.00761}_{-0.0117}$ & 
$0.1351^{+0.00769}_{-0.0112}$ & $0.1347^{+0.0105}_{-0.0115}$ & 
$0.1343^{+0.00719}_{-0.012}$\\
$100\theta_{MC}$ & 
$1.04^{+0.000578}_{-0.000573}$ & 
$1.04^{+0.00057}_{-0.000577}$ & $1.04^{+0.000585}_{-0.000588}$ & 
$1.04^{+0.000574}_{-0.000576}$\\
$\tau$ & 
$0.05395^{+0.0073}_{-0.0074}$ & 
$0.05408^{+0.00737}_{-0.00743}$ & $0.05535^{+0.00762}_{-0.00758}$ & 
$0.05483^{+0.00736}_{-0.00813}$\\
${\rm{ln}}(10^{10}A_s)$ & 
$3.044^{+0.0146}_{-0.0162}$ & 
$3.045^{+0.0154}_{-0.0153}$ & $3.045^{+0.0156}_{-0.0156}$ & 
$3.045^{+0.0154}_{-0.0172}$\\
$n_s$ &
$0.9642^{+0.00416}_{-0.00421}$ & 
$0.9633^{+0.00414}_{-0.00414}$ & $0.9658^{+0.00386}_{-0.00422}$ & 
$0.9651^{+0.00396}_{-0.00387}$\\
$w$ & 
$-1.125^{+0.0536}_{-0.0481}$ & 
$-1.102^{+0.0391}_{-0.0395}$ & $-1.089^{+0.0449}_{-0.0381}$ & 
$-1.088^{+0.0426}_{-0.0348}$\\
$\xi$ & 
$0.04985^{+0.0163}_{-0.0498}$ & 
$0.04961^{+0.0172}_{-0.0496}$ & $0.05077^{+0.0177}_{-0.0508}$ & 
$0.04866^{+0.017}_{-0.0487}$\\
\hline
$H_0$ & 
$69.47^{+0.93}_{-0.947}$ & 
$68.61^{+0.248}_{-0.248}$ & $68.74^{+0.605}_{-0.647}$ & 
$68.53^{+0.227}_{-0.225}$\\
$\Omega_{de}$ & 
$0.6723^{+0.0235}_{-0.0199}$ & 
$0.6643^{+0.0245}_{-0.0161}$ & $0.6661^{+0.0229}_{-0.0199}$ & 
$0.665^{+0.025}_{-0.0159}$\\
$\Omega_m$ & 
$0.3277^{+0.0199}_{-0.0235}$ & 
$0.3357^{+0.0161}_{-0.0245}$ & $0.3339^{+0.0199}_{-0.0229}$ & 
$0.335^{+0.0159}_{-0.025}$\\
$\sigma_8$ & 
$0.7703^{+0.0381}_{-0.0429}$ & 
$0.7673^{+0.0403}_{-0.0399}$ & $0.7589^{+0.0346}_{-0.0437}$ & 
$0.7629^{+0.0362}_{-0.0399}$\\
${\rm{Age}}/{\rm{Gyr}}$ & 
$13.76^{+0.0217}_{-0.0214}$ & 
$13.78^{+0.013}_{-0.0129}$ & $13.77^{+0.0187}_{-0.0188}$ & 
$13.77^{+0.0138}_{-0.0127}$\\
 \hline
\end{tabular}
\end{center}
\caption{Model IIA - Cosmological Parameters for different + GW Opt.}
\label{ModelIIADatasetsGWOpt}
\end{table}

\subsection{Model IIB}
\begin{figure}[!ht]
\centering
\begin{center}
\hspace{3.0cm}
\includegraphics[totalheight=3.5in]{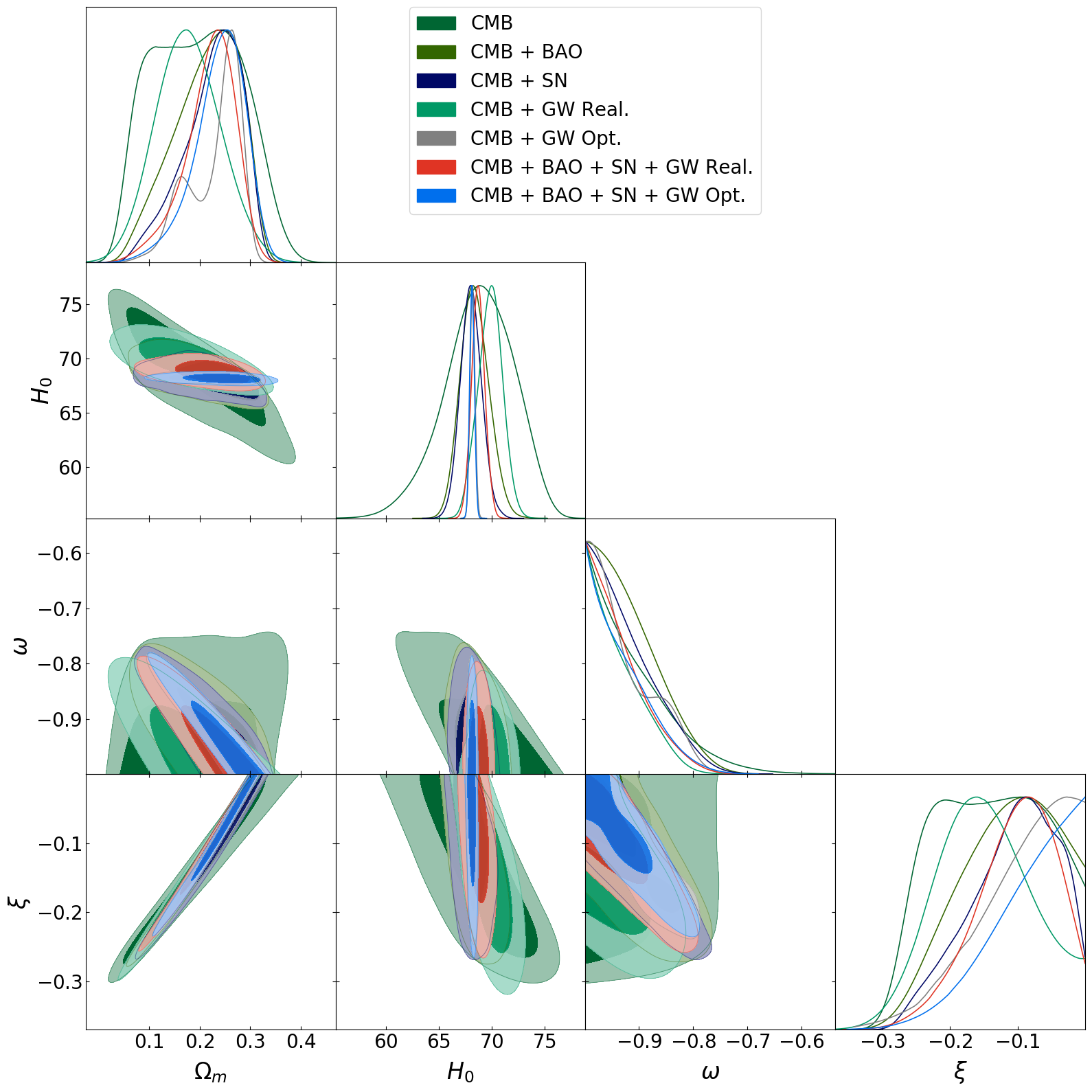}
\caption{\footnotesize  1D and 2D confidence contours for IM2B. }
\label{contoursIM2BGW}
\end{center}
\end{figure}

\hspace{0.5 cm}
 In Table \ref{ModelIIBDatasetsGWRl} and \ref{ModelIIBDatasetsGWOpt}, we list the average and $68\%$ C.L. for the parameters of Model IIB for ``+ GW Real." and the comparison between the two GW catalogs. Fig.~\ref{contoursIM2BGW} shows 1-D and 2-D confidence contours. For the coupling constant $\xi$, CMB+GW Real. provide an improvement to CMB data only ($\sim 28\%$) better than CMB+BAO ($\sim 18\%$), and equal to CMB+SN. For $H_0$, GW presents a very restrictive power, with improvements in relation to CMB data of $\sim 88\%$ to CMB+GW Real., which is better than CMB+BAO  ($\sim 85\%$) and CMB+SN ($\sim 86\%$). With respect to $\Omega_m$, CMB+GW Real. have a more restrictive power ($\sim 38\%)$ than CMB+BAO ($\sim 31\%)$ and less than CMB+SN ($\sim 41\%$), while for EoS $\omega$ is $\sim 28\%$ in comparison to CMB+BAO ($\sim 4\%$) and SN ($\sim 10\%$). All those data combined improve the constraints with respect to Planck by ($\sim 34\% $) for the interaction parameter, ($\sim 81\%$) for the Hubble constant, ($\sim 49\%$) for the matter parameter and ($\sim 26\%$) for the EoS.
 
 Comparing CMB+GW Opt. with CM+GW Real. and CMB+BAO+SN+GW Opt. with respect to CMB+BAO+SN+GW Real., we have improvements of $\sim 4\%$ and $\sim 6\%$ in the coupling and $\sim 80\%$ and $\sim 64\%$ in the Hubble constant, respectively. There is no significant improvements in $\Omega_m$ and $\omega$ due to addition of GW Opt. catalog with respect to GW Real.

\begin{table}
\footnotesize
\begin{center}
\begin{tabular}{|c|r|r|r|r|r|}
\cline{2-6}
\multicolumn{1}{c}{}  & \multicolumn{1}{|c|}{\footnotesize{ CMB}} & \multicolumn{1}{|c|}{\footnotesize{CMB + BAO}} & \multicolumn{1}{|c|}{\footnotesize{CMB + SN}} & \multicolumn{1}{|c|}{\scriptsize{CMB + GW Rl}.} & 
\multicolumn{1}{|c|}{\tiny{CMB+BAO+SN+GW Rl.}} \\ 
 \hline
 
Parameter & Avg. $\pm$ 68\% limits & Avg. $\pm$ 68\% limits & Avg. $\pm$ 68\% limits  & Avg. $\pm$ 68\% limits & Avg. $\pm$ 68\% limits \\  
 \hline
$\Omega_b h^2$ & $0.02233^{+0.000148}_{-0.000149}$ & $0.02237^{+0.000147}_{-0.000144}$ & $0.02233^{+0.000148}_{-0.000148}$ & $0.02234^{+0.000155}_{-0.000155}$ & 
$0.0224^{+0.000137}_{-0.00014}$ \\ 
$\Omega_c h^2$ & $0.06604^{+0.0452}_{-0.0275}$ & $0.07653^{+0.037}_{-0.0169}$ & $0.08125^{+0.0333}_{-0.0129}$ & $0.06287^{+0.0256}_{-0.0275}$ & 
$0.08124^{+0.0267}_{-0.0163}$ \\ 
$100\theta_{MC}$ & $1.044^{+0.00155}_{-0.00327}$ & $1.044^{+0.00102}_{-0.00242}$ & $1.043^{+0.000818}_{-0.00214}$ & $1.045^{+0.00187}_{-0.00219}$ & 
$1.043^{+0.000841}_{-0.00178}$ \\ 
$\tau$ & $0.05484^{+0.00715}_{-0.00806}$ & $0.0553^{+0.0074}_{-0.00814}$ & $0.05441^{+0.00726}_{-0.00802}$ & $0.05451^{+0.00744}_{-0.00755}$ & 
$0.05573^{+0.00747}_{-0.00803}$ \\ 
${\rm{ln}}(10^{10}A_s)$ & $3.046^{+0.0154}_{-0.0165}$ & $3.046^{+0.0154}_{-0.0167}$ & $3.045^{+0.0156}_{-0.0156}$ & $3.045^{+0.0156}_{-0.0156}$ & 
$3.046^{+0.0159}_{-0.0159}$ \\ 
$n_s$ & $0.9641^{+0.00438}_{-0.00445}$ & $0.9653^{+0.00421}_{-0.00426}$ & $0.964^{+0.00428}_{-0.00421}$ & $0.9642^{+0.00447}_{-0.00457}$ & 
$0.9661^{+0.00399}_{-0.004}$ \\ 
$w$ & $-0.9121^{+0.0216}_{-0.0869}$ & $-0.9174^{+0.022}_{-0.0816}$ & $-0.9225^{+0.0201}_{-0.0765}$ & $-0.9382^{+0.0169}_{-0.0608}$ & 
$-0.9344^{+0.015}_{-0.0646}$ \\ 
$\xi$ & $-0.1422^{+0.0886}_{-0.0903}$ & $-0.1166^{+0.0998}_{-0.0463}$ & $-0.108^{+0.0891}_{-0.0389}$ & $-0.1517^{+0.0611}_{-0.0662}$ & 
$-0.1048^{+0.0748}_{-0.0416}$ \\ 
\hline
$H_0$ & $68.84^{+3.59}_{-2.88}$ & $68.43^{+1.23}_{-1.43}$ & $68.08^{+0.999}_{-0.995}$ 
& $69.88^{+1.2}_{-1.1}$ & 
$68.74^{+0.618}_{-0.611}$ \\ 
$\Omega_{de}$ & $0.8059^{+0.101}_{-0.0862}$ & $0.786^{+0.0474}_{-0.0816}$ & $0.7744^{+0.0369}_{-0.0732}$ & $0.8229^{+0.0593}_{-0.0565}$ & 
$0.779^{+0.0371}_{-0.0577}$ \\ 
$\Omega_m$ & $0.1941^{+0.0862}_{-0.101}$ & $0.214^{+0.0816}_{-0.0474}$ & $0.2256^{+0.0732}_{-0.0369}$ & $0.1771^{+0.0643}_{-0.0755}$ & 
$0.221^{+0.0577}_{-0.0371}$ \\ 
$\sigma_8$ & $1.514^{+0.0887}_{-0.749}$ & $1.243^{+0.0294}_{-0.445}$ & $1.162^{+0.0188}_{-0.347}$ & $1.471^{+0.229}_{-0.591}$ & 
$1.145^{+0.0526}_{-0.307}$ \\ 
${\rm{Age}}/{\rm{Gyr}}$ & $13.78^{+0.052}_{-0.0788}$ & $13.78^{+0.0306}_{-0.0302}$ & $13.79^{+0.027}_{-0.027}$ & $13.75^{+0.0238}_{-0.0236}$ & 
$13.77^{+0.0182}_{-0.0182}$ \\ 
 \hline
\end{tabular}
\end{center}
\caption{Model IIB - Cosmological Parameters for different datasets + GW Real.}
\label{ModelIIBDatasetsGWRl}
\end{table}

\begin{table}
\footnotesize
\begin{center}
\begin{tabular}{|c|r|r|r|r|r|}
\cline{2-5}
\multicolumn{1}{c}{}  & 
\multicolumn{1}{|c|}{\scriptsize{CMB + GW Rl}.} & 
\multicolumn{1}{|c|}{\scriptsize{CMB + GW Opt.}} & 
\multicolumn{1}{|c|}{\tiny{CMB+BAO+SN+GW Rl.}} & \multicolumn{1}{|c|}{\tiny{CMB+BAO+SN+GW Opt.}} \\
 \hline
 Parameter & Avg. $\pm$ 68\% limits & Avg. $\pm$ 68\% limits & Avg. $\pm$ 68\% limits  & Avg. $\pm$ 68\% limits \\ 
 \hline
$\Omega_b h^2$ & 
$0.02234^{+0.000155}_{-0.000155}$ & 
$0.0224^{+0.000137}_{-0.000137}$ & $0.0224^{+0.000137}_{-0.00014}$ & 
$0.02242^{+0.000138}_{-0.000136}$\\
$\Omega_c h^2$ & 
$0.06287^{+0.0256}_{-0.0275}$ & 
$0.08431^{+0.0296}_{-0.0351}$ & %
$0.08124^{+0.0267}_{-0.0163}$ & 
$0.08856^{+0.0265}_{-0.0169}$\\
$100\theta_{MC}$ & 
$1.045^{+0.00187}_{-0.00219}$ & 
$1.043^{+0.00228}_{-0.00194}$ & $1.043^{+0.000841}_{-0.00178}$ & 
$1.043^{+0.000916}_{-0.00168}$\\
$\tau$ & 
$0.05451^{+0.00744}_{-0.00755}$ & 
$0.05562^{+0.00778}_{-0.00781}$ & $0.05573^{+0.00747}_{-0.00803}$ & 
$0.05611^{+0.0073}_{-0.00829}$\\
${\rm{ln}}(10^{10}A_s)$ & 
$3.045^{+0.0156}_{-0.0156}$ & 
$3.046^{+0.0161}_{-0.016}$ & $3.046^{+0.0159}_{-0.0159}$ & 
$3.046^{+0.0154}_{-0.017}$\\
$n_s$ & 
$0.9642^{+0.00447}_{-0.00457}$ & 
$0.9657^{+0.00404}_{-0.00402}$ & $0.9661^{+0.00399}_{-0.004}$ & 
$0.9668^{+0.00392}_{-0.00396}$\\
$w$ & 
$-0.9382^{+0.0169}_{-0.0608}$ & 
$-0.926^{+0.0206}_{-0.073}$ & $-0.9344^{+0.015}_{-0.0646}$ & 
$-0.9308^{+0.0205}_{-0.0682}$\\
$\xi$ & 
$-0.1517^{+0.0611}_{-0.0662}$ & 
$-0.09746^{+0.0975}_{-0.0246}$ & $-0.1048^{+0.0748}_{-0.0416}$ & 
$-0.08489^{+0.0849}_{-0.0236}$\\
\hline
$H_0$ & 
$69.88^{+1.2}_{-1.1}$ & 
$68.21^{+0.228}_{-0.231}$ & $68.74^{+0.618}_{-0.611}$ & 
$68.16^{+0.218}_{-0.22}$\\
$\Omega_{de}$ & 
$0.8229^{+0.0593}_{-0.0565}$ & 
$0.7692^{+0.0755}_{-0.0643}$ & $0.779^{+0.0371}_{-0.0577}$ & 
$0.7597^{+0.0384}_{-0.0572}$\\
$\Omega_m$ & 
$0.1771^{+0.0565}_{-0.0593}$ & 
$0.2308^{+0.0643}_{-0.0755}$ & $0.221^{+0.0577}_{-0.0371}$ & 
$0.2403^{+0.0572}_{-0.0384}$\\
$\sigma_8$ & 
$1.471^{+0.229}_{-0.591}$ & 
$1.103^{+0.0443}_{-0.296}$ & $0.2308^{+0.0565}_{-0.0593}$ & 
$1.053^{+0.0078}_{-0.38}$\\
${\rm{Age}}/{\rm{Gyr}}$ & 
$13.75^{+0.0238}_{-0.0236}$ & 
$13.78^{+0.013}_{-0.0129}$ & $13.77^{+0.0182}_{-0.0182}$ & 
$13.78^{+0.0129}_{-0.0129}$\\
 \hline
\end{tabular}
\end{center}
\caption{Model IIB - Cosmological Parameters for GW Real. and GW Opt.}
\label{ModelIIBDatasetsGWOpt}
\end{table}

\subsection{Model III}
\hspace{0.5 cm}
As we can see in Table~\ref{ModelIIIDatasetsGWRl}, there is an improvement of $\sim 35\%$ in the coupling constant due to addition of GW Real. catalog to CMB data. This improvement is better than CMB+SN ($\sim 12\%$), but inferior to CMB+BAO ($\sim 54\%$). For the Hubble constant, CMB+ GW Real. shows an improvement of $\sim 87\%$ with respect to CMB data only, while CMB+BAO and CMB+SN show an improvement of $\sim 85\%$ and $\sim 86\%$, respectively. With respect to $\Omega_m$, CMB+GW Real. have a more restrictive power ($\sim 81\%)$ than CMB+BAO ($\sim 76\%)$ and CMB+SN ($\sim 55\%$), while for EoS $\omega$ is $\sim 63\%$ in comparison to CMB+BAO ($\sim 81\%$) and SN  ($\sim 79\%$). The improvements with respect to Planck data only using all data together are: ($\sim 63\%$) for $\xi$, ($\sim 94\%$) for $H_0$, ($\sim 88\%$) for $\Omega_m$ and ($\sim 90\%$) for $\omega$.

\begin{figure}[!ht]
\centering
\begin{center}
\hspace{3.0cm}
\includegraphics[totalheight=3.5in]{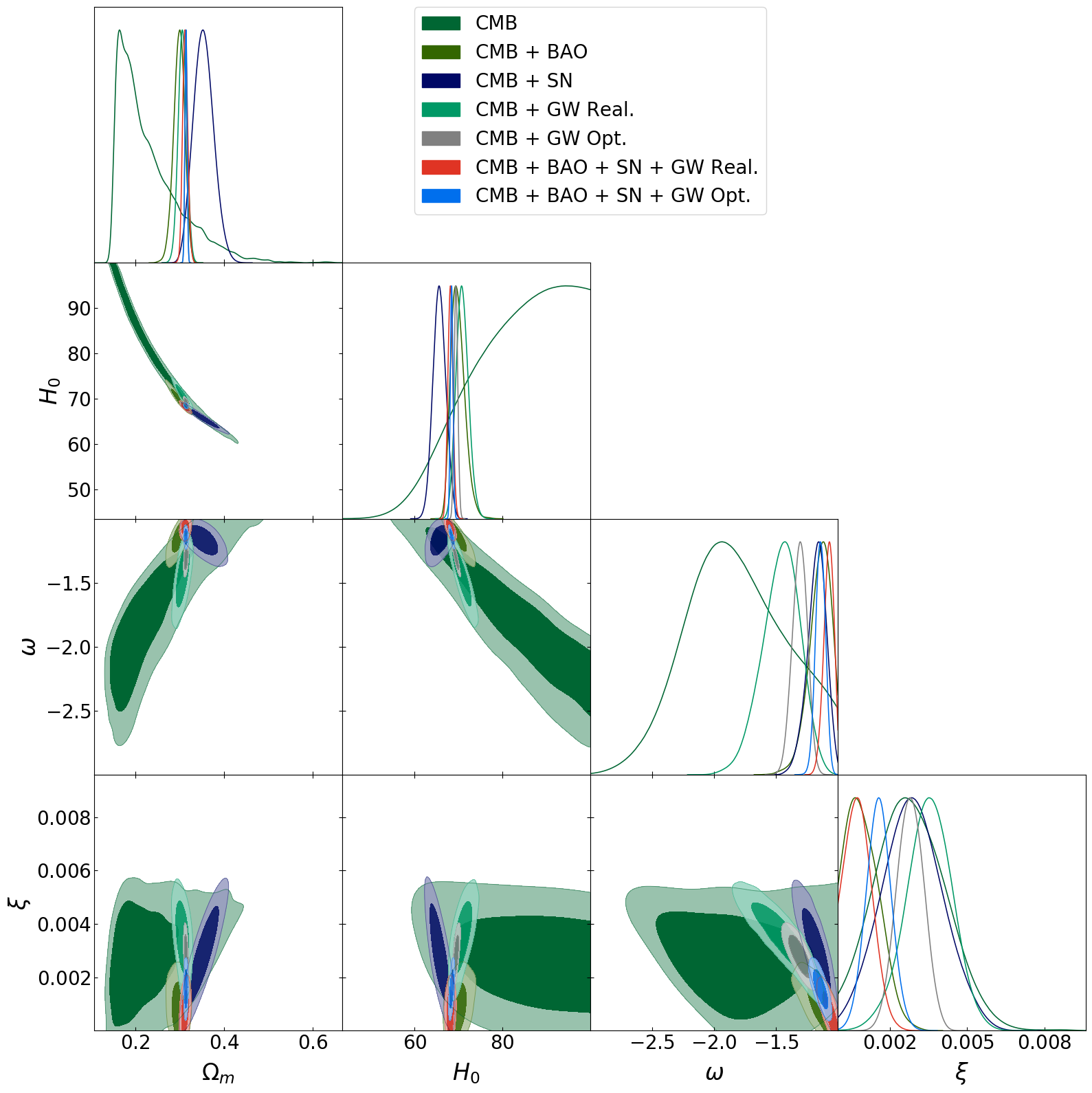}
\caption{\footnotesize 1D and 2D confidence contours for IM3. }
\label{contoursIM3GW}
\end{center}
\end{figure}

Using CMB+GW Opt., there is a decrease of $\sim 39\%$ in the coupling constant error and $\sim 70\%$ in the Hubble constant error with respect to CM+GW Real., as can be seen in Table~\ref{ModelIIIDatasetsGWOpt}. In the same table, we can see that there is an improvement of only $\sim 4\%$ in $\xi$ between CMB+BAO+SN+GW Opt. and CMB+BAO+SN+GW Real., and an improvement $\sim 47\%$ in $H_0$. For $\Omega_m$, they are respectively $\sim 74\%$ and $\sim 59\%$, and for $\omega$ they are respectively $\sim 60\%$ and $\sim 2\%$. The 1-D and 2-D contours are shown in Fig.~\ref{contoursIM3GW}.

\begin{table}
\footnotesize
\begin{center}
\begin{tabular}{|c|r|r|r|r|r|}
\cline{2-6}
\multicolumn{1}{c}{}  & \multicolumn{1}{|c|}{\footnotesize{ CMB}} & \multicolumn{1}{|c|}{\footnotesize{CMB + BAO}} & \multicolumn{1}{|c|}{\footnotesize{CMB + SN}} & \multicolumn{1}{|c|}{\scriptsize{CMB + GW Rl}.} & 
\multicolumn{1}{|c|}{\tiny{CMB+BAO+SN+GW Rl.}} \\ 
 \hline
 
Parameter & Avg. $\pm$ 68\% limits & Avg. $\pm$ 68\% limits & Avg. $\pm$ 68\% limits  & Avg. $\pm$ 68\% limits & Avg. $\pm$ 68\% limits \\  
 \hline
$\Omega_b h^2$ & $0.02261^{+0.000176}_{-0.0002}$ & $0.02249^{+0.000161}_{-0.000163}$ & $0.02257^{+0.000181}_{-0.000183}$ & $0.02264^{+0.00018}_{-0.000179}$ & 
$0.02245^{+0.000161}_{-0.000158}$ \\ 
$\Omega_c h^2$ & $0.128^{+0.00365}_{-0.00429}$ & $0.1224^{+0.00172}_{-0.002}$ & $0.1288^{+0.00352}_{-0.00356}$ & $0.1303^{+0.00257}_{-0.00226}$ & 
$0.1219^{+0.00135}_{-0.00147}$ \\ 
$100\theta_{MC}$ & $1.04^{+0.000422}_{-0.00038}$ & $1.041^{+0.000312}_{-0.000313}$ & $1.04^{+0.000363}_{-0.000353}$ & $1.04^{+0.000323}_{-0.000319}$ & 
$1.041^{+0.00029}_{-0.000298}$ \\ 
$\tau$ & $0.05146^{+0.00735}_{-0.00742}$ & $0.05464^{+0.0075}_{-0.00755}$ & $0.05189^{+0.0072}_{-0.00714}$ & $0.05045^{+0.00722}_{-0.00723}$ & 
$0.05456^{+0.00766}_{-0.00759}$ \\ 
${\rm{ln}}(10^{10}A_s)$ & $3.033^{+0.016}_{-0.0161}$ & $3.042^{+0.0159}_{-0.0161}$ & $3.035^{+0.0152}_{-0.0151}$ & $3.03^{+0.0153}_{-0.0151}$ & 
$3.042^{+0.0163}_{-0.0157}$ \\ 
$n_s$ & $0.96^{+0.005}_{-0.00491}$ & $0.9646^{+0.00408}_{-0.00403}$ & $0.9593^{+0.00445}_{-0.00476}$ & $0.9583^{+0.00423}_{-0.00426}$ & 
$0.9648^{+0.004}_{-0.00393}$ \\ 
$w$ & $-1.822^{+0.379}_{-0.445}$ & $-1.151^{+0.11}_{-0.0596}$ & $-1.173^{+0.0859}_{-0.0641}$ & $-1.461^{+0.17}_{-0.134}$ & 
$-1.077^{+0.047}_{-0.0325}$ \\ 
$\xi$ & $0.002846^{+0.00126}_{-0.00147}$ & $0.001005^{+0.000382}_{-0.000871}$ & $0.002889^{+0.00119}_{-0.00119}$ & $0.003521^{+0.000887}_{-0.000875}$ & 
$0.0008411^{+0.000374}_{-0.000635}$ \\ 
\hline
$H_0$ & $83.35^{+16.6}_{-5.08}$ & $69.9^{+1.25}_{-1.85}$ & $65.64^{+1.46}_{-1.45}$ & $70.87^{+1.31}_{-1.5}$ & 
$68.14^{+0.556}_{-0.572}$ \\ 
$\Omega_{de}$ & $0.7691^{+0.0814}_{-0.0213}$ & $0.7018^{+0.0112}_{-0.0134}$ & $0.6464^{+0.024}_{-0.0215}$ & $0.6941^{+0.00938}_{-0.00971}$ & 
$0.6877^{+0.00607}_{-0.00543}$ \\ 
$\Omega_m$ & $0.2309^{+0.0213}_{-0.0814}$ & $0.2982^{+0.0134}_{-0.0112}$ & $0.3536^{+0.0215}_{-0.024}$ & $0.3059^{+0.00971}_{-0.00938}$ & 
$0.3123^{+0.00543}_{-0.00607}$ \\ 
$\sigma_8$ & $0.8485^{+0.31}_{-0.855}$ & $0.8344^{+0.021}_{-0.0193}$ & $0.7568^{+0.225}_{-0.804}$ & $0.7687^{+0.17}_{-0.81}$ & 
$0.8199^{+0.012}_{-0.0112}$ \\ 
${\rm{Age}}/{\rm{Gyr}}$ & $13.83^{+0.139}_{-0.207}$ & $13.81^{+0.0391}_{-0.0463}$ & $14.03^{+0.107}_{-0.108}$ & $14.^{+0.0616}_{-0.06}$ & 
$13.84^{+0.029}_{-0.0369}$ \\ 
 \hline
\end{tabular}
\end{center}
\caption{Model III - Cosmological Parameters for different datasets + GW Real.}
\label{ModelIIIDatasetsGWRl}
\end{table}

\begin{table}
\footnotesize
\begin{center}
\begin{tabular}{|c|r|r|r|r|r|}
\cline{2-5}
\multicolumn{1}{c}{}  & 
\multicolumn{1}{|c|}{\scriptsize{CMB + GW Rl}.} & 
\multicolumn{1}{|c|}{\scriptsize{CMB + GW Opt.}} & 
\multicolumn{1}{|c|}{\tiny{CMB+BAO+SN+GW Rl.}} & 
\multicolumn{1}{|c|}{\tiny{CMB+BAO+SN+GW Opt.}} \\
 \hline
 
 Parameter & Avg. $\pm$ 68\% limits & Avg. $\pm$ 68\% limits & Avg. $\pm$ 68\% limits  & Avg. $\pm$ 68\% limits \\
 \hline
$\Omega_b h^2$ & 
$0.02264^{+0.00018}_{-0.000179}$ & 
$0.02258^{+0.000166}_{-0.000165}$ & $0.02245^{+0.000161}_{-0.000158}$ & 
$0.02249^{+0.000165}_{-0.000168}$\\
$\Omega_c h^2$ & 
$0.1303^{+0.00257}_{-0.00226}$ & 
$0.1282^{+0.00138}_{-0.00137}$ & $0.1219^{+0.00135}_{-0.00147}$ & 
$0.1239^{+0.00124}_{-0.00123}$\\
$100\theta_{MC}$ & 
$1.04^{+0.000323}_{-0.000319}$ & 
$1.04^{+0.0003}_{-0.000295}$ & $1.041^{+0.00029}_{-0.000298}$ & 
$1.041^{+0.000289}_{-0.00029}$\\
$\tau$ & 
$0.05045^{+0.00722}_{-0.00723}$ & 
$0.05134^{+0.00739}_{-0.0074}$ & $0.05456^{+0.00766}_{-0.00759}$ & 
$0.05427^{+0.00718}_{-0.00791}$\\
${\rm{ln}}(10^{10}A_s)$ & 
$3.03^{+0.0153}_{-0.0151}$ & 
$3.033^{+0.0158}_{-0.0156}$ & $3.042^{+0.0163}_{-0.0157}$ & 
$3.04^{+0.0156}_{-0.0158}$\\
$n_s$ & 
$0.9583^{+0.00423}_{-0.00426}$ & 
$0.9597^{+0.00401}_{-0.00399}$ & $0.9648^{+0.004}_{-0.00393}$ & 
$0.9636^{+0.00394}_{-0.00388}$\\
$w$ & 
$-1.461^{+0.17}_{-0.134}$ & 
$-1.309^{+0.0634}_{-0.0572}$ & $-1.077^{+0.047}_{-0.0325}$ & 
$-1.142^{+0.0393}_{-0.0386}$\\
$\xi$ & 
$0.003521^{+0.000887}_{-0.000875}$ & 
$0.002796^{+0.000535}_{-0.000535}$ & $0.0008411^{+0.000374}_{-0.000635}$ & 
$0.001573^{+0.00048}_{-0.000483}$\\
\hline
$H_0$ & 
$70.87^{+1.31}_{-1.5}$ & 
$69.43^{+0.39}_{-0.431}$ & 
$68.14^{+0.556}_{-0.572}$ & 
$68.44^{+0.293}_{-0.298}$\\
$\Omega_{de}$ & 
$0.6941^{+0.00938}_{-0.00971}$ & 
$0.6859^{+0.00241}_{-0.00245}$ & $0.6877^{+0.00607}_{-0.00543}$ & 
$0.686^{+0.00235}_{-0.00235}$\\
$\Omega_m$ & 
$0.3059^{+0.00971}_{-0.00938}$ & 
$0.3141^{+0.00245}_{-0.00241}$ & $0.3123^{+0.00543}_{-0.00607}$ & 
$0.314^{+0.00235}_{-0.00235}$\\
$\sigma_8$ & 
$0.7687^{+0.17}_{-0.81}$ & 
$0.7429^{+0.251}_{-0.792}$ & $0.8199^{+0.012}_{-0.0112}$ & 
$0.8245^{+0.0141}_{-0.00868}$\\
${\rm{Age}}/{\rm{Gyr}}$ & 
$14.^{+0.0616}_{-0.06}$ & 
$13.96^{+0.0325}_{-0.0325}$ & $13.84^{+0.029}_{-0.0369}$ & 
$13.88^{+0.027}_{-0.0274}$\\
 \hline
\end{tabular}
\end{center}
\caption{Model III - Cosmological Parameters for GW Real. and GW Opt.}
\label{ModelIIIDatasetsGWOpt}
\end{table}

\newpage
\section{Conclusions}
\hspace{0.5 cm}
Gravitational waves as standard sirens can be a very useful cosmological probe in the near future. Third generation detectors like Einstein Telescope can improve current GWs observations and have sensibility to detect an order of $10^2$ events per year, which is enough to impose constraints as good as the current cosmological probes.

We consider a non trivial dark sector where dark matter and dark energy interact with each other. Assuming three phenomenological interacting models as fiducial cosmologies, we generate a ``realistic" and ``optimistic" joint GW+GRB mock catalog and use them as standard sirens to forecast possible constraints in these models. In general, as we increase the number of events, we obtain more restrictive confidence contours. We also see that the most sensitive parameter due to standard sirens is the Hubble constant. 

The addition of simulated gravitational wave standard sirens to current CMB data provide constraints as good or better than the addition of BAO and SN to CMB data. For the Hubble constant there is a decrease of almost $\sim 90\%$ in error in relation to CMB data only, which suggests that standard sirens can help solving the tension in $H_0$ in the near future. It has also been shown that the addition of GW data to CMB data can help to break the degeneracy between the parameters. 


Our results are compatible with the results given in Yang \textit{et al}. \cite{Yang:2019bpr} for interacting vacuum-energy models, where the coupling has the form $Q=3H\xi\rho_{\unit{de}}$ as in our model II but the  DE equation of state is fixed, $w=-1$. The authors of \cite{Yang:2019bpr} found an improvement of $17\%$ in $\xi$ and $35\%$ in $H_0$ due to the addition of GW simulated data to CMB+BAO+SN data. In another work, Yang \textit{et al}. found that an addition of GW SS to CMB data from \textit{Planck} can reduce the uncertainty on the DM-DE coupling $\xi$ by a factor of 5 \cite{Yang:2019vni}. They analyzed an interacting dark energy model with a coupling equivalent to our Model I. All these works have provided evidence of the great power that can be reached using standard sirens, improving the constraints obtained by current cosmological probes. Concluding, gravitational waves will be a very useful observable in cosmology. 

 For possible future work, we plan to investigate modified gravitational wave propagation \cite{Belgacem:2017ihm, Belgacem:2018lbp, Belgacem:2019pkk, Belgacem:2019tbw} due to the presence of an interaction in dark sector.

\section{Acknowledgements}
\hspace{0.5 cm}

This work was supported by CAPES (Coordena\c{c}\~{a}o de Aperfei\c{c}oamento de Pessoal de N\'{i}vel Superior) and FAPESP (Funda\c{c}\~{a}o de Amparo \`{a} Pesquisa do Estado de S\~{a}o Paulo). We truly thanks an anonymous referee, whose suggestions greatly improved this work.

\bibliographystyle{JHEP.bst}
\bibliography{JCAPSS.bbl}

\end{document}